\documentclass[12pt]{iopart}
\usepackage{epsfig}
\usepackage{color}
  
\begin{document}    

\title{Discontinuous phase transition in an open-ended Naming Game} 
\author{Nuno Crokidakis $^{\ddag}$ and Edgardo Brigatti $^{\star}$}  
  
\address{$^{\ddag}$ Instituto de F\'{\i}sica, Universidade Federal Fluminense \\
Av. Litor\^anea s/n,
24210-340, Niter\'oi, RJ, Brazil}
\address{$^{\star}$ Instituto de F\'{\i}sica, Universidade Federal do Rio de Janeiro \\
Av. Athos da Silveira Ramos 149,
Cidade Universit\'aria, 21941-972, Rio de Janeiro, RJ, Brazil}
\ead{nuno@if.uff.br, edgardo@if.ufrj.br}

\date{\today}
  
\begin{abstract}
\noindent
In this work we study on a 2-dimensional square lattice a recent version of the Naming Game, an agent-based model used for describing the emergence of linguistic structures. The system is  open-ended and agents can invent new words all along the evolution of the game, picking them up from a pool characterised by a Gaussian distribution with standard deviation $\sigma$. The model displays a nonequilibrium phase transition at a critical point $\sigma_{c}\approx 25.6$, which separates an absorbing consensus state from an active fragmented state where agents continuously exchange different words. The finite-size scaling analysis of our simulations strongly suggests that the phase transition is discontinuous.

\end{abstract}
    

\maketitle


\section{Introduction}

\qquad
In the last years, an original model for describing 
spreading of conventions, generally denominated Naming Game 
\cite{baronchelli06},
has attracted the attention of the scientific community 
interested in modelling linguistic related phenomena.
Despite - or perhaps because of - its simplicity,  
agent-based models directly inspired by 
the Naming Game, have shown enough power 
to unravel the key factors involved in such phenomena as 
the birth of neologisms \cite{baronchelli08},
the self-organisation of a hierarchical category structure in a population 
of individuals \cite{puglisi08},
the emergence of universality in colour naming patterns \cite{baronchelli10}, 
the rise of a protosyntactic structure \cite{brigatti12} and the so-called duality 
of patterning in human communication \cite{tria12}.

The key ingredients present in this model, which make it
well suited to cope with the description of linguistic 
phenomena, can be roughly summarised 
in quite realistic but simples rules of interaction 
conjugated with the ability
of describing a flexible and rich individual memory.

In the Naming Game the interaction is build up on a
reinforcement rule which fixes
a linguistic convention throughout
the selective exclusion of unshared variants. This process  is a memory-based negotiation strategy, made up of a sequence of trials that shape and reshape the system memories. Memory and feedbacks are naturally encompassed 
by these rules which result to be more widespread than 
simple imitation mechanisms that can be described by 
Ising-like dynamics. 

The memory of the different linguistic variants 
is guaranteed by an individual inventory
which can contain an infinite number of symbols.
In fact, it is structured by 
an array of potentially infinite cells where each 
cell is set on one of 
an infinite number of possible numerable states.
It follows that the number of possible states
an agent can experience
is limited only by the dynamics of symbols invention and exchange.
This is a really important fact, considering that
arbitrariness and variation is a structural characteristic of linguistic units
which forces toward a model scheme which must
encompass the open-ended nature of the linguistic system \cite{sauss}. 
For this reason, agents must be characterised by a memory capable 
of accumulating an unlimited number of possible traits, each one 
characterised by an infinite number of possible choices.

In the original model , the problem of limiting this 
variety, necessary for reaching consensus,
was solved  by allowing the introduction 
of different words just in the first time steps.
In this first stage of the simulation 
the initial random conditions
in terms of variant diversity are fixed.
In the following of the dynamics, words diffuse throughout agreement and learning mechanisms.
Learning allows the memorisation of
a new word in the case of a failed communication, 
causing a fast increase in the overlap between the words 
exchanged by the community.

Recently variants of this 
implementation of the Naming Game \cite{brigatti08, brigatti09}
explored the possibility of 
a constant introduction of new words 
with the intention of 
defining a really open-ended system,
where convention creation is always a possibility, 
and not reduced to the definition of the initial conditions.
Anyway, the negotiation dynamics is able to 
order the system toward consensus only if 
some type of restriction on the freedom of invention 
is introduced.
In fact, also in nature, the linguistic conventions are not totally arbitrary:  
the continuity with the history of the linguistic community 
restricts the freedom of invention \cite{sauss}, 
limiting the open-ended character of symbols production. 
These considerations suggest to describe the reservoir of words 
by means of a probability distribution which models the presence 
of some constraints, differentiating between more and less 
probable words.

The introduction of these different ingredients \cite{brigatti09},
which at a first sight can seem 
a minor modification of the classical implementation,
are quite interesting for two reasons.
First, because this variant of the model is able to explore the nature and effects of
ordering in an open-end system.
Second, because it naturally introduces
a nontrivial 
phase transition between a consensus absorbing state,
characterised by the use of 
the same single world by all the players,
and an active stationary state of
disordered and fragmented clusters, 
where agents continuously exchange different words.
This transition is driven by the tuning of a control parameter
which limits the degree of variance of the invented words.
A similar transition is obtained also in the original Naming Game,
at the expense of introducing a noise term.
In that work \cite{baronchelli07}, a parameter modulates 
the level of noise present in communications (no noise corresponds to the 
original Naming Game),
generating a limitation in the efficiency of the negotiation dynamics.
For a sufficiently low level of noise, consensus is reached.

The principal purpose of this work is to study and characterise the mentioned transition, throughout an accurate numerical analysis. The first work in this direction \cite{brigatti09} presented such transition for 
the model implemented on the fully-connected graph.
Here, we look at the model on a 2-dimensional square lattice, 
with the purpose of giving a description of the critical behaviour
with the help of Monte Carlo simulations and finite-size scaling analysis. 
In other words, we are interested in analysing the impact of short-range interactions on the dynamics of the model and on the phase transition that occurs between the above-mentioned states.

Nonequilibrium phase transitions in ordering dynamics 
associated with the description of opinion formation
have attracted increasing theoretical interest among 
the community of the statistical physicists \cite{loreto_review}.
Even in the absence of a general theory for nonequilibrium critical behaviour,
their characterisation in terms of different universality classes and the definition
of the prominent and robust ones, is a very important issue  \cite{odor04}.
In this classification effort, because of the lack of theoretical basis, 
it is not obvious which parameters are relevant.
Inspired by the equilibrium systems
and based on general grounds, loosely speaking
one expects to have to pay attention to the dimensionality of the system, 
the number of absorbing states, the symmetry of the problem
and in some cases to the details of the dynamics, 
with the hope of this last aspect being relevant only in 
accidental cases, rather than in generic.

Considering systems embedded in a two-dimensional space
common results exist for models which 
display the equilibrium Ising class \cite{deoliveira93}. 
Another widespread class is 
the Directed Percolation universality class 
for some models which exhibit a single absorbing state
or double absorbing states 
with some asymmetry in the dynamics which favours one of the two possible absorbing states 
\cite{dickman,privman,hinrichsen}. 
Finally, associated with critical phenomena for systems with two symmetric absorbing states, recently appeared the Generalised Voter universality 
class \cite{dornic01, castellano12},
which seems to be an important and solid class,
characterised by the logarithmic decay of the density of interfaces
and others characteristic critical exponents.

Clearly, our system has fundamental 
different ingredients from the ones listed above.
In particular, a proliferation of possible absorbing
states appears with a lack of symmetry in the dynamics.
The presence of a large number of possible 
final absorbing states is a trait commonly observed in
 the Axelrod' s  model for the formation of cultural domains \cite{axelrod}.
In such model, the nature of the consensus-fragmentation
phase transition turns from continuous to discontinuous, 
increasing the number of cultural features above two \cite{castellano00}. 
This behaviour also occurs in other 
models with more than two symmetric absorbing states 
that display a discontinuous phase transition \cite{lipowski02}.
Also in our system, the transition between the ordered consensus state 
and the disordered phase with fragmented clusters 
characterised by heterogeneous memories 
clearly seems to be a discontinuous transition.

In the following we will describe the details of the model (Section 2),
the behaviour of the dynamics on a 2-dimensional lattice
compared with the implementation on the fully-connected graph (Section 3), 
and finally the numerical analysis for the characterisation of the phase transition (Section 4).


\section{The model}

\qquad
We simulate a naming game played by $P$ agents 
embedded on a regular 2-dimensional square lattice with 
$P=L\times L$ sites and periodic boundary conditions.
This implementation defines  a short-range-interaction system,
where agents can interact only with their four nearest neighbours.

In the initial state each player starts with an empty inventory. 
At each time step, a pair of agents is randomly selected 
and their actions are governed by the following
rules, which are the same used in the former implementation
on the fully-connected graph \cite{brigatti09}:

1) The first agent (speaker) selects one of his/her
 words or, if his/her inventory is empty, he/she creates a new one. 
 After that, the word is transmitted to the second agent (hearer).

2a) If the hearer possesses the transmitted word, the communication is a success. 
The two agents update their inventories so as to keep only the word involved in the interaction. 

2b) Otherwise, the communication is a failure. 
The speaker invents a new word which must be
different from all the other ones present in his/her
 inventory.\\

New words are invented in the early stages of the simulation,
when inventories are empty, or at anytime, if the communication 
is a failure. This second option is not present in the 
original Naming Game \cite{baronchelli06}, where, in the case of a failed 
communication, the hearer learns the speaker's selected word.

Invention is performed picking up 
a word from a pool of different words which are sorted out
with a probability characterised by a
Gaussian distribution centred in zero with standard deviation 
$\sigma$.
This distribution was chosen 
because of the efficiency of its 
numerical implementation. 
Words are determined taking the integer part of the 
absolute value of the sorted real number.
We can observe that the reservoir must be unlimited for allowing that 
any invented new word can be always different from the ones present in the speaker's memory.
Anyway, it must be characterised by a distribution which falls to zero very quickly, 
like the Gaussian one, otherwise consensus can be hardly  
attained, as can be seen for the implementation realised on the fully-connected graph \cite{brigatti09}.\\


\section{Dynamical behaviour}

\begin{figure}[h]
\begin{center}
\includegraphics[width=0.5\textwidth, angle=0]{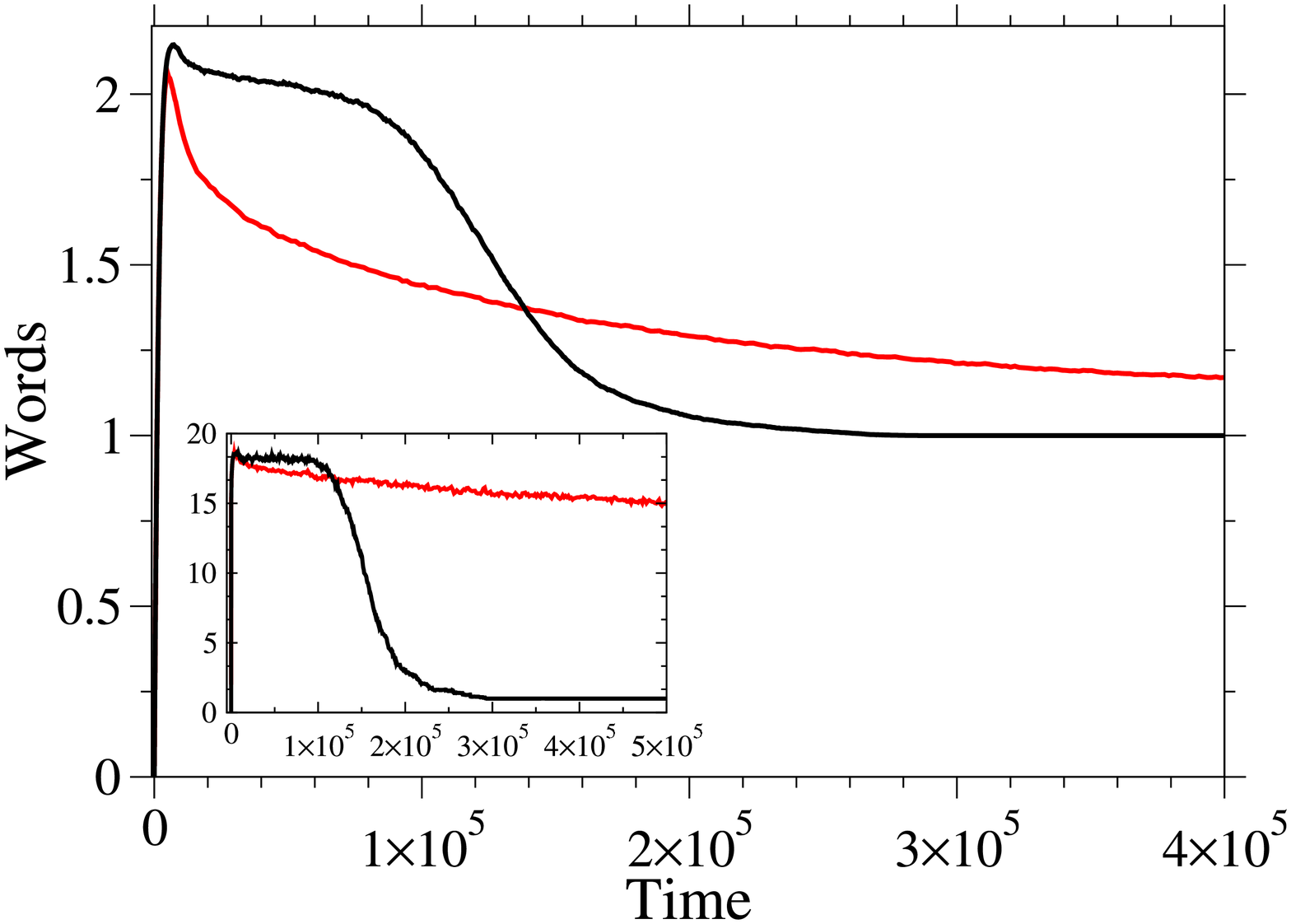}
\vspace{0.9cm}
\includegraphics[width=0.5\textwidth, angle=0]{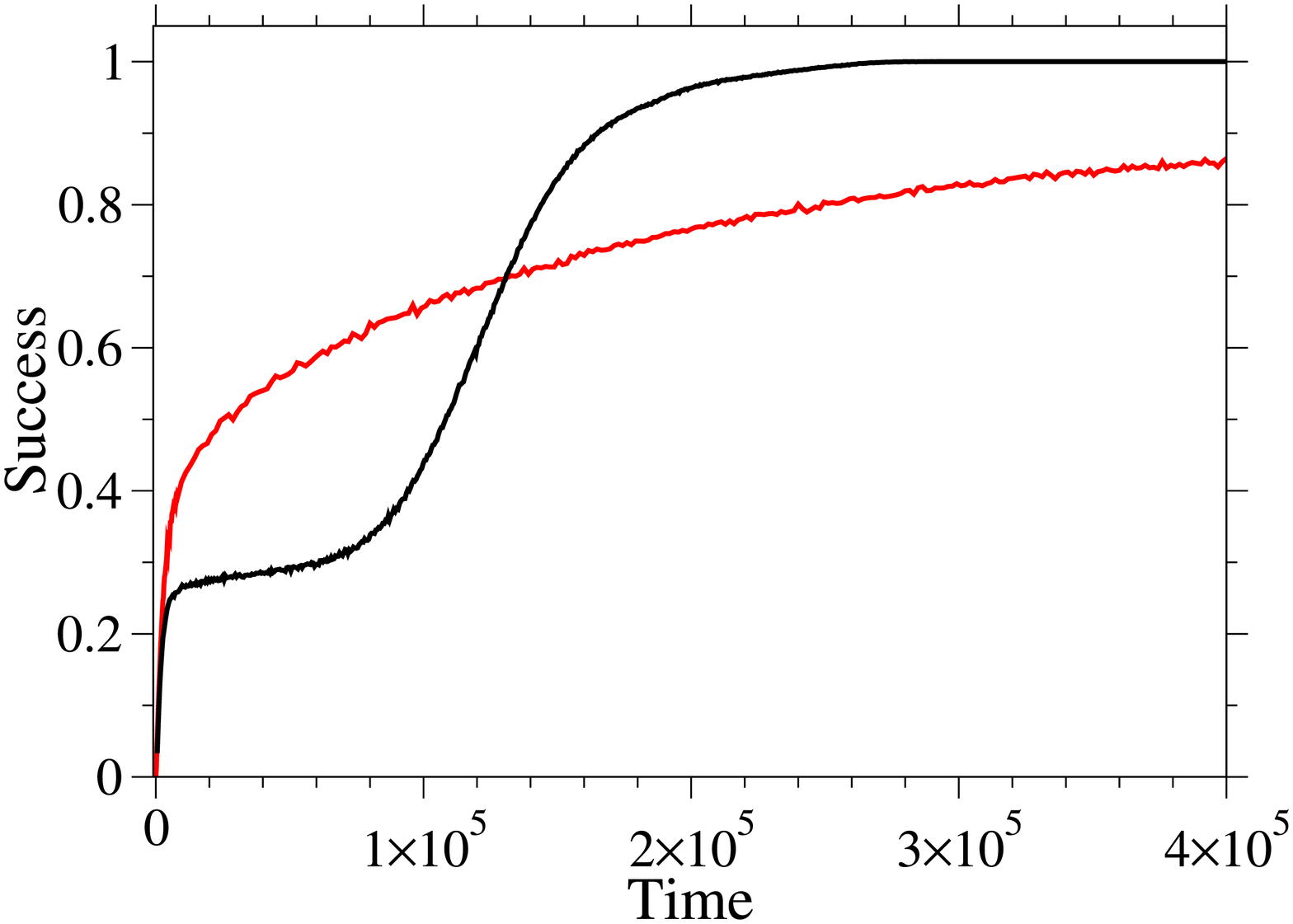}
\end{center}
\caption{\small Time evolution in the fully-connected graph (black) and in the 2-dimensional lattice model (red). The convergence of the simulations on the lattice is reached only after more than $1.4\times10^6$ time steps. \textit{Top}: temporal evolution for the total number of words divided by the total population. In the inset, the maximum number of different words. \textit{Bottom}: the success rate $S(t)$. Data are averaged over 100 simulations, $P=1600$, $\sigma=5$. }
\label{fig_pheno}
\end{figure}

\qquad
In the following we unfold a comparison between the typical results of the fully-connected graph implementation \cite{brigatti09}, with the ones obtained in the 2-dimensional lattice (see Figure~\ref{fig_pheno}).

We describe the time evolution of our system on the basis
of some usual global quantities \cite{baronchelli06}, namely the total number 
of words ($N_{t}(t)$), the number of different words ($N_{d}(t)$) 
present in the population and the success rate ($S(t)$), 
which measures the average rate of success of communications. 
This is obtained by evaluating the frequency of successful 
communications in a given time interval.
In Figure \ref{fig_pheno} we can appreciate the time evolution 
of these quantities for the lattice implementation compared with the 
fully-connected graph one. 
In both situations $N_{t}$ shows a
fast initial transient, during which
many different words are created.
In fact, agents start with an empty inventory and, 
in each interaction, each speaker invents at least 
a new word and each hearer can possibly learn one.
After this early stage,  a maximum is reached.
In the fully-connected graph implementation 
this stage is followed by a long plateau slightly decreasing along the time evolution.
In contrast, the lattice simulations, after having crossed
the maximum, abruptly 
diminish the number of total words, directly
starting a long way towards the ordered phase.
These different behaviours can be easily understand.
On the fully-connected graph each agent can interact with all other players. 
This fact  increases the possibility of unsuccessful interactions 
which determine the invention of new words. During this phase,
a success in communication is rare and words are hardly eliminated,
generating the observed long and declining plateau.
Only when the redundancy of words reaches a sufficient high level
the system undergoes the ordering transition.
The number of successful games increases 
and $N_{t}$ changes its concavity and begins a decay 
towards the consensus state, corresponding to 
one common word for all the players. 
In contrast, on the lattice, 
players only communicate locally with the same four neighbours.
This means that words spread locally and, 
in general,  the number of success in
communications  is significant and agents have a 
limited necessity to invent new words.
It follows that the maximum value reached by the memory 
is smaller and no plateau is observed, since 
the coarsening of the single-state clusters
soon reduces the number of words.
However, in this condition, the final convergence 
to the consensus state is much slower.
These considerations are supported by and coherent 
with the dynamics displayed by $N_{d}(t)$ and $S(t)$ (see Figure \ref{fig_pheno}).

\begin{figure}[t]
\begin{center}
\includegraphics[width=0.5\textwidth, angle=0]{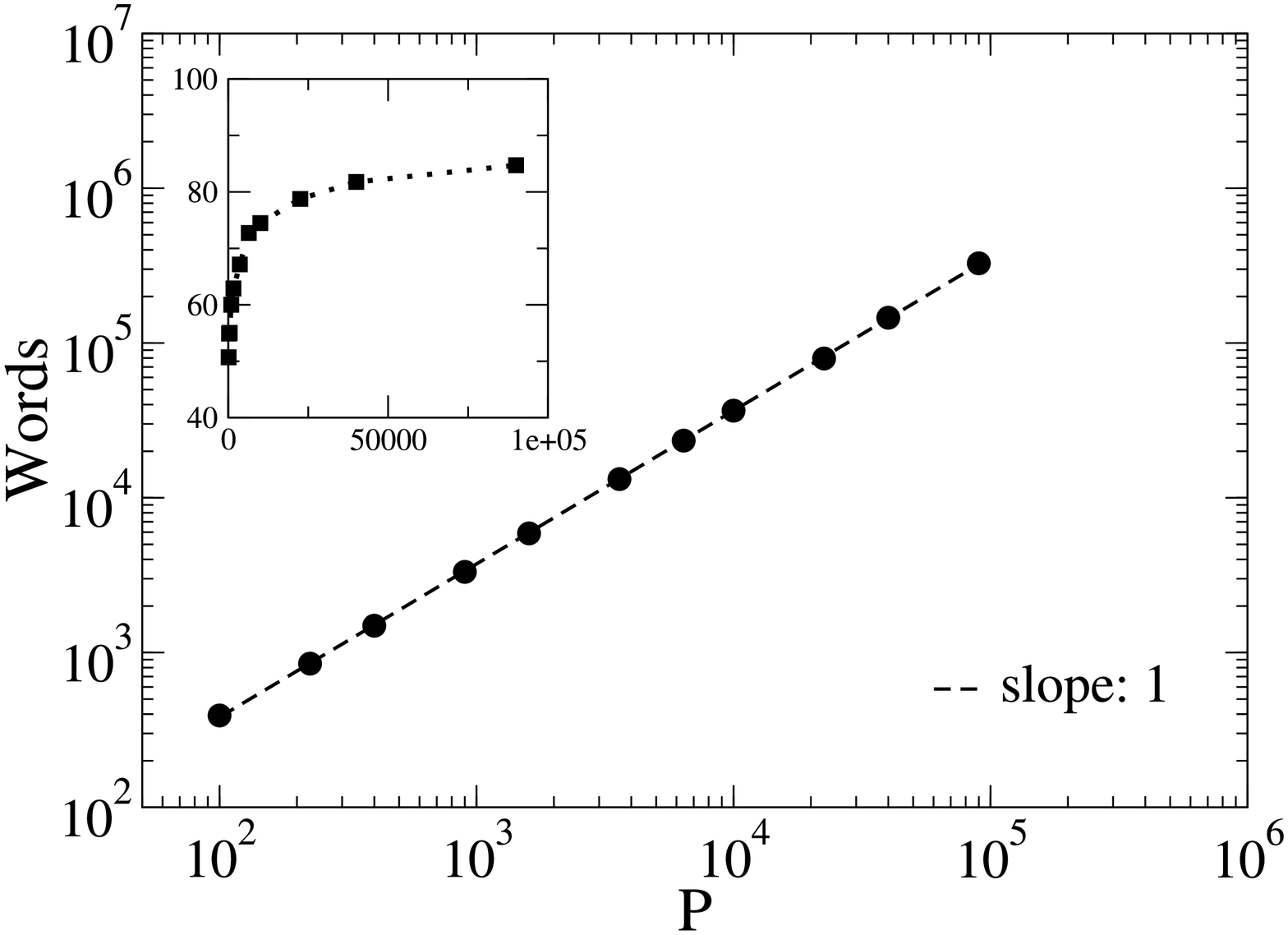}
\includegraphics[width=0.5\textwidth, angle=0]{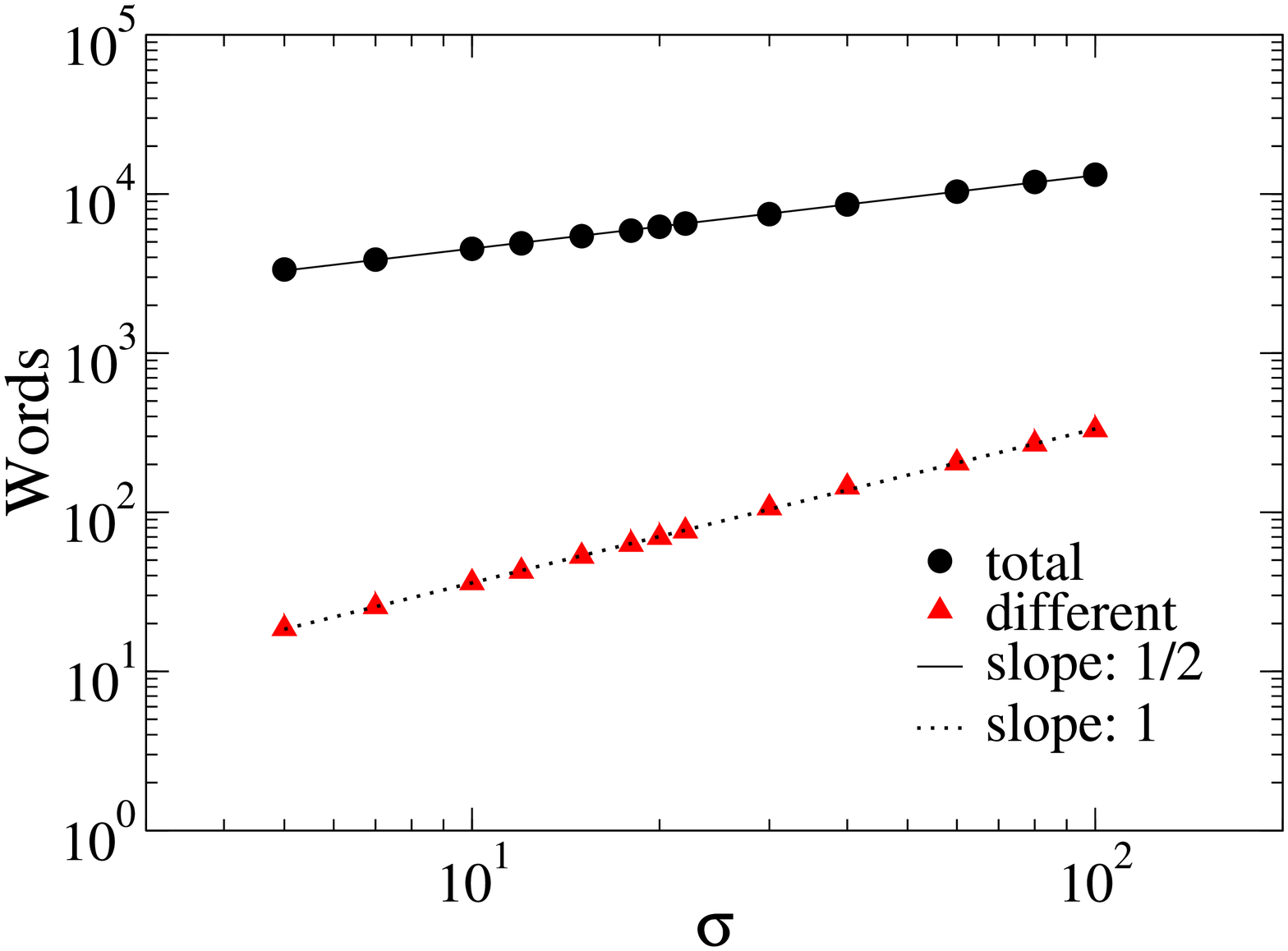}
\end{center}
\caption{\small \textit{Top}: Maximum number of total words (the dashed line has slope $1$) and, in the inset, maximum number of different words for different population sizes ($\sigma=18$). \textit{Bottom}: Maximum number of total words (circles) and maximum number of different words (triangles) for different $\sigma$ values ($L=40$). Data are averaged over $100$ simulations. 
}
\label{fig_dynamics}
\end{figure}
 
Figure \ref{fig_dynamics} describes the scaling behaviour
of the system memory size. The maximum number of 
total words linearly scales with the population size $P=L^{2}$.
So, like in the mean-field case, the number of total words of each player 
is not dependent on the community size \cite{brigatti09}.
The fact that the same behaviour is maintained is quite obvious,
considering that we can not have a dependence slower than linear.
The same scaling law was observed in the standard Naming Game implemented on a 2-dimensional lattice, where the total memory scales as P \cite{baronchelliR}.

We also studied the scaling behaviour in dependence of the $\sigma$ 
value: $N_{d}$ follows a linear scaling and  $N_{t} \sim \sigma^{1/2}$. 
In summary, no modification of the scaling behaviour of the mean-field case \cite{brigatti09} is observed.

\begin{figure}[h]
\begin{center}
\vspace{0.4cm}
\includegraphics[width=0.5\textwidth, angle=0]{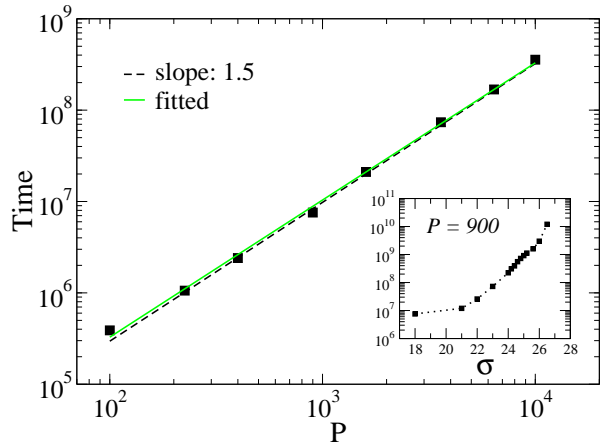}
\end{center}
\caption{\small 
Convergence time for different population sizes P ($\sigma=18$). On the fully-connected graph a linear dependence in P was found \cite{brigatti09}. In the inset it is exhibited the convergence time for a fixed population ($P=900$) as a function of $\sigma$. Data are averaged over 100 simulations.
}
\label{fig_tcritic}
\end{figure}

Finally, we explored the behaviour of the convergence time $T_c$ 
(the average time needed to reach consensus),
which is clearly a quantity of particular interest in the context
of the social dynamics applications.
We analysed the dependence of $T_c$ on P obtaining 
$T_{c} \sim P^{1.5}$ (see the main plot of Figure \ref{fig_tcritic}).
As expected, the time needed to reach consensus
grows in relation to the fully-connected graph case, 
where, to a first approximation, $T_c$ scales linearly with $P$. 
It is interesting to remember that
the original Naming Game simulated on 
a 2-dimensional lattice has a convergence time equal to $P^{2}$,
with an exponent 0.5 larger than the fully-connected graph case \cite{baronchelliR}, as it is found in our implementation.

In the inset of Figure \ref{fig_tcritic}, we report $T_{c}$ as a function of the parameter $\sigma$ for a fixed population size ($L=30$). For low $\sigma$ values, $T_{c}$ increases exponentially and, for higher values, it points towards a divergence, signalling the fact that for large $\sigma$ convergence is not attained. A similar behaviour was observed in the fully-connected graph case, and is associated with the proximity of the phase transition point. This transition will be analysed in more details in the next section.


\section{Phase transition}

\qquad
In this section we study the stationary  
states of the system with the aim of analysing 
the transition properties of the model.
The phase transition is characterised 
by the passage
from a consensus absorbing state,
characterised by the use of 
the same single world by all the players,
to an active stationary state of
disordered and fragmented clusters, 
where agents continuously interchange different words.
Consensus is reached only if the variance 
of the distribution of the invented words is sufficiently small.
In this case, a unique cluster characterised by the same
word emerges. 
In contrast, for large $\sigma$, convergence
is not attained and small clusters with one or more
different words are present in the system.
In this regime the invention of words outperforms the capacity 
of the population to reach consensus because the variety
of the new words is too hight and generates
too much diversity among the players.

Inspired by previous studies \cite{castellano12,castellano00}, we decided on 
the relative size of the largest cluster $(C_{m}/L^{2})$ for the role of the  
order parameter of the model.
We define  $C_{m}$ as the size of the largest cluster 
composed by agents sharing the same unique word
and we measure it at the end of each simulation.
In addition, we estimate the fluctuations (or the susceptibility) 
$\chi$ of this order parameter and its fourth-order cumulant $U$, which can be obtained from the simulations by
evaluating

\begin{eqnarray}
\chi & = & L^{2}(<C_{m}^{2}> - <C_{m}>^{2}) \\
U & = & 1 - \frac{<C_{m}^{4}>}{3<C_{m}^{2}>^{2}} ~,
\end{eqnarray}

where $<\;\;>$ stands for averages over different simulations taken at the steady states. We generate simulations with L ranging from 20 up to 60. For these L values the system presents a remarkable slow relaxation time close to the transition, which obliges us to run very long simulations which last $4 \times 10^{10}$  Monte Carlo steps. 

\begin{figure}[h]
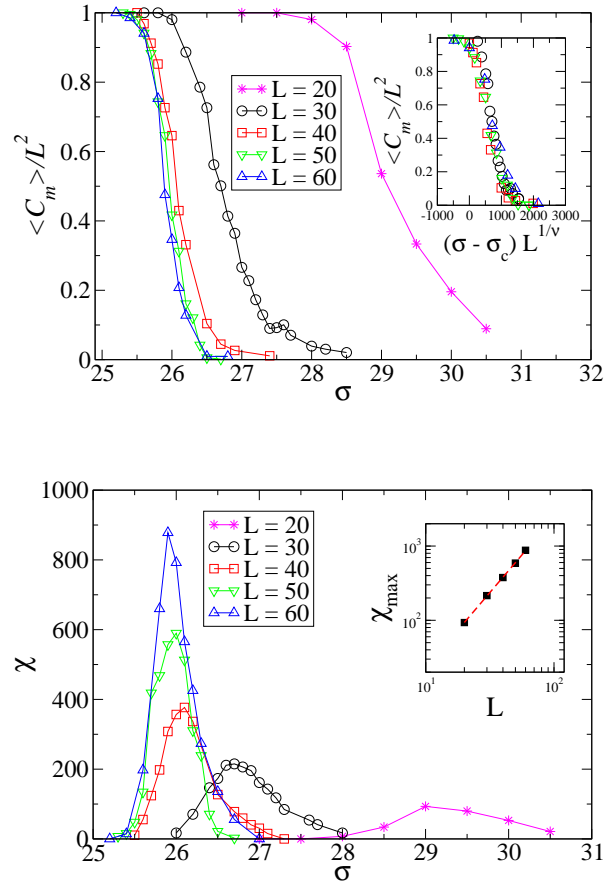

\begin{center}
\vspace{0.6cm}
\includegraphics[width=0.5\textwidth, angle=0]{Cm_x_sd.eps}
\\
\vspace{1.0cm}
\includegraphics[width=0.5\textwidth, angle=0]{flucts_x_sd.eps}
\end{center}
\caption{\small \textit{Top}: Mean largest cluster size $<C_m>$ normalised over the number of total sites as a function of $\sigma$ for some lattice sizes $L$ (main plot). In the inset it is exhibited the corresponding scaling plot. The best collapse of data was obtained for a critical point $\sigma_{c}\approx 25.6$ and for $1/\nu\approx 1.95$, which corresponds to the system dimension $d=2$ within error bars. \textit{Bottom}: fluctuations $\chi$ of $C_{m}/L^2$ (or susceptibility, in the main plot) as a function of $\sigma$ for some lattice sizes $L$ and the corresponding values of the peaks $\chi_{{\rm max}}$ for each size $L$ (inset). From the fitting of these last data, we obtained a straight line with slope $\gamma/\nu\approx 2.02$, which corresponds to the system dimension $d=2$ within error bars. All results are averaged over $100$ samples.}
\label{fig_phtrans}
\end{figure}

In Figure~\ref{fig_phtrans} we exhibit the behaviour of the order parameter (top) and its fluctuations (bottom) as functions of $\sigma$ for some lattice sizes $L$. One can observe the typical behaviour of a phase transition, i.e., the order parameter goes to zero at critical values $\sigma_{c}(L)$ and the susceptibility peaks $\chi_{{\rm max}}$ grow for increasing values of $L$. Following the standard procedure for finite-size scaling (FSS) analysis \cite{loreto_review,dickman,hinrichsen,nuno}, one can consider that the above-mentioned quantities present power-law scalings near the transition point $\sigma_{c}$ of the form

\begin{eqnarray}
<C_{m}>/&L^{2}&  \sim  L^{-\beta/\nu} ~, \\
&\chi & \sim  L^{\gamma/\nu} ~, \\
&U & \sim  b\,L^{a} ,
\end{eqnarray}
where $\beta$, $\gamma$, $\nu$, $a$ are critical exponents and $b$ is a constant. In continuous phase transitions these exponents present values that satisfy the relation $2\,\beta + \gamma = d\,\nu$, where $d$ is the dimension of the system, $a=0$ and $b>0$. In the case of discontinuous (first-order) phase transitions, we have $\beta=0$, $\nu=1/d$, $\gamma/\nu=d$, $a=d$ and $b<0$ \cite{odor04,fisher,binder,salinas,nuno,nuno2}. In other words, for discontinuous transitions the susceptibility peaks diverge to $+\infty$ as $\chi_{{\rm max}} \sim L^{d}$, whereas the fourth-order cumulant presents a well-defined minimum that diverges to $-\infty$ as $U_{{\rm min}} \sim -L^{d}$. 
\begin{figure}[t]
\begin{center}
\vspace{0.5cm}
\includegraphics[width=0.5\textwidth, angle=0]{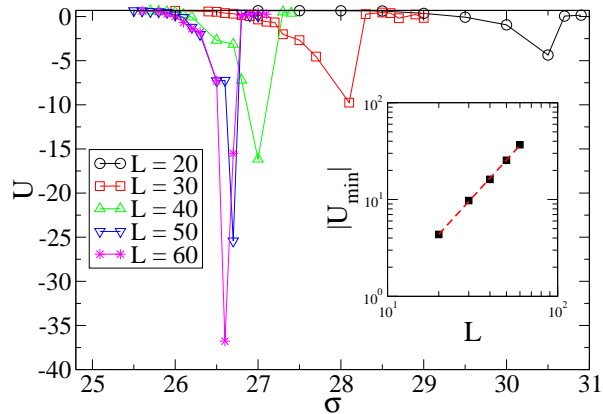}
\end{center}
\caption{\small 
Binder cumulant as a function of $\sigma$ for typical lattice sizes $L$, showing the characteristic well-defined minimum near the transition. In the inset we exhibit the modulus of the values of these minima as a function of $L$. Fitting data, we obtain a straight line with slope $\approx 1.97$, which corresponds to the system dimension $d=2$ within error bars.
}
\label{fig_binder}
\end{figure}

Taking into account the above FSS relations, in the insets of Figure~\ref{fig_phtrans} we exhibit the scaling plot of the order parameter and the susceptibility peaks as functions of $L$ in the log-log scale. The former is obtained for $\sigma_{c}\approx 25.6$ and $1/\nu\approx 1.95$, which give us an estimate of the critical point $\sigma_{c}=\sigma_{c}(L\to\infty)$. In addition, we obtain the results $\beta/\nu=0$ and $1/\nu\approx d$, which suggests a discontinuous transition. The value of $\chi_{{\rm max}}$ grows with the system size with $\gamma/\nu\approx 2.02$, which corresponds to the system dimension $d=2$ within error bars. All these results strongly suggest the occurrence of a discontinuous transition. 
Furthermore, as can be seen in Figure~\ref{fig_binder}, the fourth-order cumulant presents well-defined minima near the transition. Looking for the minima values, the modulus $|U_{{\rm min}}|$ grows with an exponent $a\approx 1.97$, which is also compatible with a discontinuous transition ($a\approx d$) \cite{salinas,nuno2}.

\begin{figure}[h]
\begin{center}
\includegraphics[width=0.5\textwidth, angle=0]{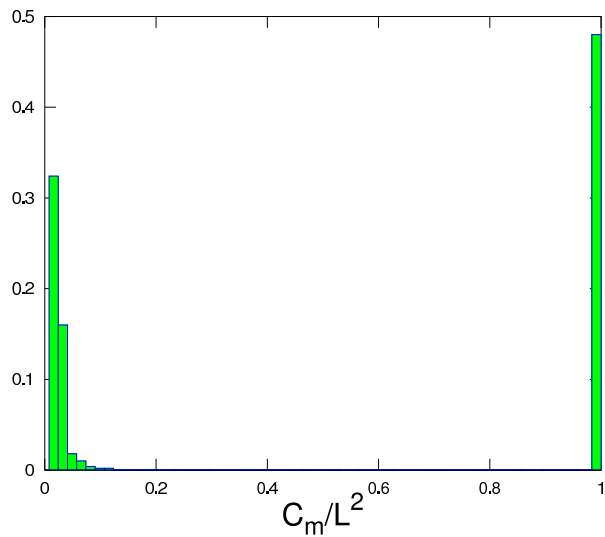}
\end{center}
\caption{\small 
Probability distribution function of the normalised largest cluster size. 
A clear bimodal distribution appear near $\sigma_{c}(L)$.
Incidentally, we can observe that, in general, if consensus is not attained, 
single-state clusters sharing the same word are small.
Data are obtained from the final state of $500$
different simulations running with $\sigma=26.7$ and $L=30$. 
}
\label{fig_distribution}
\end{figure}

Finally, we plot the probability distribution function (PDF) of the order parameter near the transition. In the case of a continuous transition, this quantity presents a one-peak structure. On the other hand, for discontinuous transitions the PDF is double peaked. In Figure~\ref{fig_distribution} we show the results for $L=30$ and $\sigma=26.7$, i.e., near the transition for the mentioned size (see Figure~\ref{fig_phtrans}). In this case, we clearly observe the two-peak structure, which confirms the discontinuous character of this transition.

\section{Conclusions}

\qquad
In this work we studied on a regular 2-dimensional lattice 
the open-ended Naming Game introduced in \cite{brigatti09},
where it was implemented on the fully-connected graph.
 
In a first step we explored the differences which the
topology can generate in the dynamics of the model.
In general, we found that the implementation on a
regular lattice limits the necessity of
invention of new words causing a smaller  memory effort.
However, the scaling behaviour presents no modification from the 
mean-field  case for the maximum number of total and different words.
On the other side, the convergence to the consensus state is remarkably slower and the time needed to reach consensus grows with an exponent $0.5$ larger than the case of the fully-connected graph.

Next, we pointed our attention 
to the principal purpose of our work:
the analysis of the phase transition properties 
of the system. 
In fact, this model is characterised 
by a transition
between a consensus absorbing state,
where the same single world is used by all the players,
and an active stationary state of
fragmented clusters distinguished by
the presence of different words.
Consensus is reached if the variance of the invented words is small enough.

The presence of a double-peaked probability distribution function 
of the order parameter near the transition suggests
a  phase transition of discontinuous type.
We tested this supposition by means of a
finite-size scaling analysis.
The estimation of the $\beta$ and $\nu$ exponents,
the dependence of $\chi_{{\rm max}}$ on the system size,
and the behaviour of the fourth-order cumulant,
were all compatible with a discontinuous transition.
From this analysis we also estimated the value of the critical point 
$\sigma_{c}\approx 25.6$, which is significantly larger than
in the fully-connected graph implementation ($6<\sigma_{c}^{FC}<7$), 
a fact that can be ascribed to the minor propensity for inventing new words.

Interestingly, these results can be compared with the
transition obtained in the original Naming Game with a noise term \cite{baronchelli07}.
Also in that work, a discontinuous phase transition between 
consensus and polarised states was found and 
analytically described in the case of a limited number of opinions. 

\section*{Acknowledgments} 
NC is grateful to Universidade Federal Fluminense through the FOPESQ project for partial financial support. EB acknowledges a grant from FAPERJ (111.883/2013).

\end{document}